\documentclass[fleqn,12pt]{wlscirep}
\usepackage{aas_macros}
\usepackage{setspace}
\usepackage{subcaption}
\usepackage{caption}
\usepackage{color}

\usepackage[symbol]{footmisc}
\usepackage{lineno}
\usepackage{amsmath}
\usepackage{multirow}

\title{Discovery of oscillations above 200\,keV in a black hole X-ray binary with \emph{Insight}-HXMT}
\author[1]{Xiang Ma}
\author[1]{Lian Tao}
\author[1,2]{Shuang-Nan Zhang}
\author[1,3]{Liang Zhang}
\author[1,4]{Qing-Cui Bu}
\author[1]{Ming-Yu Ge}
\author[1]{Yu-Peng Chen}
\author[1]{Jin-Lu Qu}
\author[1]{Shu Zhang}
\author[1]{Fang-Jun Lu}
% ...

\affil[1]{Key Laboratory of Particle Astrophysics, Institute of High Energy Physics, Chinese Academy of Sciences, 19B Yuquan Road, Beijing 100049, People's Republic of China}
\affil[2]{University of Chinese Academy of Sciences, Chinese Academy of Sciences, Beijing 100049, People's Republic of China}
\affil[3]{Physics \& Astronomy, University of Southampton, Southampton, Hampshire SO17 1BJ, UK}
\affil[4]{Institut f\"{u}r Astronomie und Astrophysik, Kepler Center for Astro and Particle Physics, Eberhard Karls Universit\"{a}t, Sand 1, 72076 T\"{u}bingen, Germany}
\affil[5]{Key Laboratory for Research in Galaxies and Cosmology, Shanghai Astronomical Observatory, Chinese Academy of Sciences, 80 Nandan Road, Shanghai 200030, People's Republic of China}
\affil[6]{School of Astronomy and Space Science, University of Chinese Academy of Sciences, 100049, Beijing, People's Republic of China}
\affil[7]{Institute of High Energy Physics, Chinese Academy of Sciences, 19B Yuquan Road, Beijing 100049, People's Republic of China}
\affil[8]{Department of Astronomy, Beijing Normal University, Beijing 100088, China}
\affil[9]{Department of Physics, Tsinghua University, Beijing 100084, China}
\affil[10]{Department of Astronomy, Tsinghua University, Beijing 100084, People's Republic of China}
\affil[11]{Department of Engineering Physics, Tsinghua University, Beijing 100084, China}
\affil[12]{Computing Division, Institute of High Energy Physics, Chinese Academy of Sciences, 19B Yuquan Road, Beijing 100049, People's Republic of China}
\affil[13]{Key Laboratory of Space Astronomy and Technology, National Astronomical Observatories, Chinese Academy of Sciences, Beijing 100012}
\affil[14]{School of Physics and Optoelectronics, Xiangtan University, Yuhu District, Xiangtan, Hunan, 411105, China}
\affil[15]{College of Physics, Jilin University, No. 2699 Qianjin Street, Changchun City, 130012, China}

\begin{abstract}
{
Low-frequency quasi-periodic oscillations (LFQPOs) are commonly found in black hole X-ray binaries, and their origin is still under debate. The properties of LFQPOs at high energies (above 30\,keV) are closely related to the nature of the accretion flow in the innermost regions, and thus play a crucial role in critically testing various theoretical models. The Hard X-ray Modulation Telescope (\emph{Insight}-HXMT) is capable of detecting emissions above 30\,keV, and is therefore an ideal instrument to do so. Here we report the discovery of LFQPOs above 200 keV in the new black hole MAXI J1820+070 in the X-ray hard state, which allows us to understand the behaviours of LFQPOs at hundreds of kiloelectronvolts. The phase lag of the LFQPO is constant around zero below 30\,keV, and becomes a soft lag (that is, the high-energy photons arrive first) above 30\,keV. The soft lag gradually increases with energy and reaches $\sim$0.9\,s in the 150--200\,keV band. The detection at energies above 200\,keV, the large soft lag and the energy-related behaviors of the LFQPO pose a great challenge for most currently existing models, but suggest that the LFQPO probably originates from the precession of a small-scale jet.
}

\end{abstract}

\begin{document}
\maketitle

LFQPOs have been found in most transient black-hole X-ray binaries\cite{Motta2015}, and their frequencies usually range from a few mHz to 30\,Hz\cite{Motta2016,Ingram2019}. LFQPOs are believed to originate from the inner part of the accretion flow, however, the physical mechanism is not well understood. Theories such as instabilities in accretion flow (e.g. fluctuations in the mass accretion rate\cite{Tagger1999,Cabanac2010,Rodriguez2002,Varniere2002,Varniere2012}) or geometrical effects (e.g. precession of the inner hot flow\cite{Ingram2009,Veledina2013,Ingram2016,Ingram2017} or the jet base\cite{Stevens2016,deRuiter2019}) are proposed to explain their origins. The properties of LFQPOs above 30\,keV directly depend on the radiation mechanism and geometry in the innermost regions of the accretion flow, and are key to understand the origin of LFQPOs. Previous observations have detected rich LFQPOs below 30\,keV, but LFQPOs at much higher energies are expected in some previous studies\cite{Rodriguez2004}, though rarely reported so far. With the wide energy range of the Hard X-ray Modulation Telescope (\emph{Insight}-HXMT)\cite{Zhang2014,Zhang2020}, we may detect high-energy LFQPOs above 30\,keV in some black-hole binaries and study further into LFQPOs.

MAXI J1820+070 is a new Galactic black hole (BH)\cite{Torres2019} discovered by the Monitor of All-sky X-ray Image (MAXI) on 11 March 2018\cite{Kawamuro2018}. \emph{Insight}-HXMT carried out a Target of Opportunity (ToO) observation three days after its discovery, and has monitored the whole outburst more than 140 times with a total exposure of 2000\,ks. Thanks to the large effective area of \emph{Insight}-HXMT in the hard X-ray band ($\sim5000~\rm cm^2$ in 20--250\,keV), the peak net count rate of the source above 100 keV can be up to 150 counts~$\rm s^{-1}$. The high statistics for high-energy photons and the broad energy coverage (1--250 keV) of \emph{Insight}-HXMT allow us to do detailed timing analysis on high-energy and broad-band variability.

\begin{figure}
\centering
\includegraphics[width=0.9\textwidth]{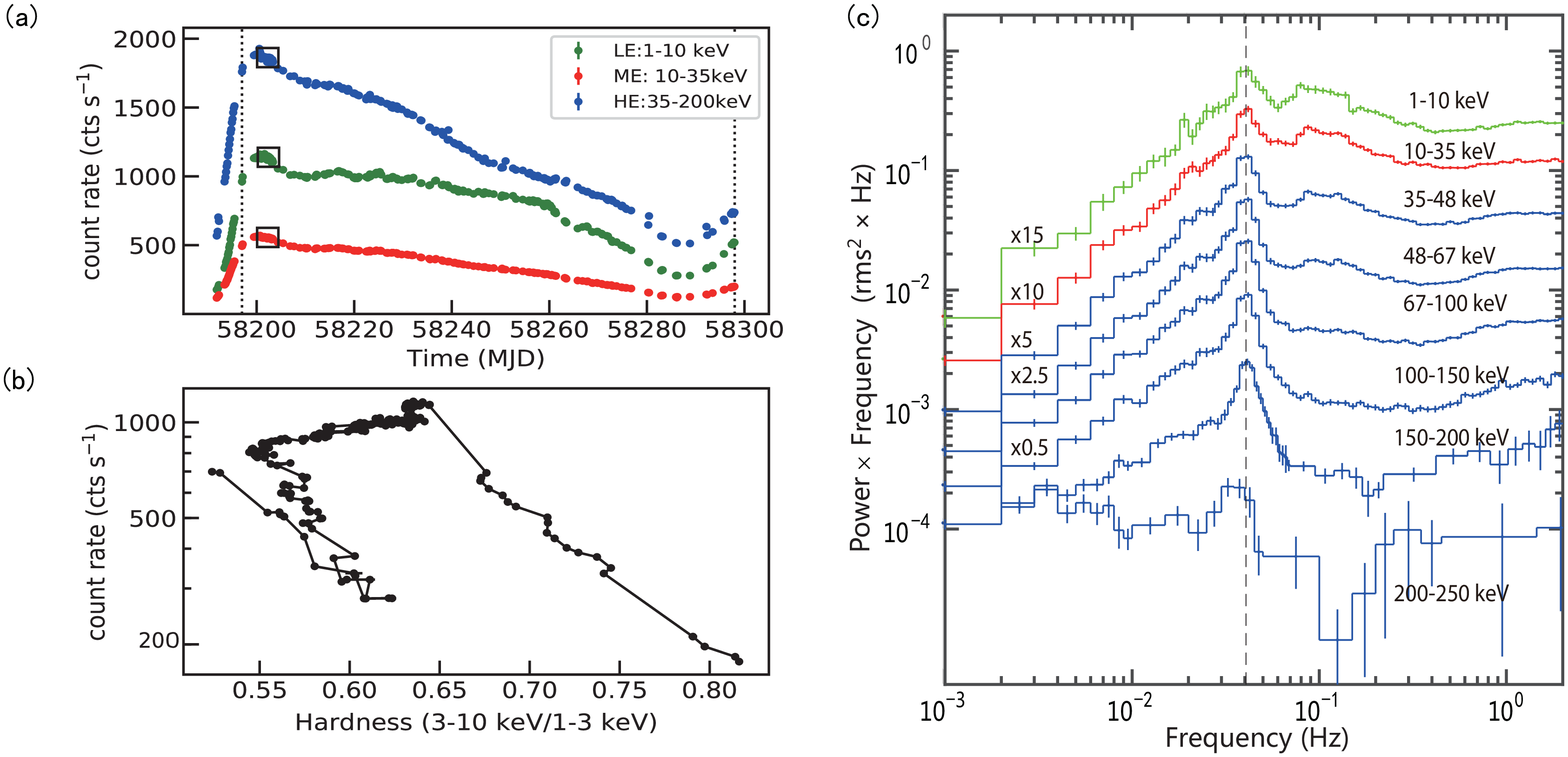}
\caption{Light curves, hardness-intensity diagram and power density spectra (PDS) of MAXI J1820+070 in the X-ray hard state. (a) \emph{Insight}-HXMT light curves of MAXI J1820+070. The two vertical dashed lines indicate the period in which the LFQPOs are detected. (b) The hardness-intensity diagram, defined as the total 1--10\,keV count rate versus the ratio of hard (3--10\,keV) to soft (1--3\,keV) count rate. (c) The PDS of different energy bands: 1--10\,keV for LE; 10--35\,keV  for ME; 35--48\,keV, 48--67\,keV, 67--100\,keV, 100--150\,keV, 150--200\,keV and 200--250\,keV for HE. We artificially multiply the power by a different factor for plotting clarity. The vertical dashed line indicates the centroid frequency of the LFQPO. The PDS of the longest observation taken on MJD 58201.3--58203.3 (ObsID P0114661004, exposure of $\sim 67$\,ks for HE) are plotted. The observation is indicated by the boxes in panel (a). Error bars in this paper correspond to 1$\sigma$ confidence intervals.}
\label{figure1}
\end{figure}

Throughout most of its X-ray hard state, where its flux variability is large and spectrum is dominated by a hard power-law component (see {\bf Methods} for the details of states and transitions), we observed an LFQPO in all three telescopes (LE: 1--10 keV; ME: 10--30 keV; HE: 25--250 keV) of \emph{Insight}-HXMT (see Figure \ref{figure1}). The quality factor ($Q$) of the LFQPO, defined as the ratio of the QPO centroid frequency ($\nu$) to its full width at half maximum ($\Delta\nu$), is larger than 2.5. The LFQPO in different energy bands has a similar centroid frequency, accompanied by a flat-top noise component dominating at lower frequencies. The detection significance for the longest observation is $\sim 4~\sigma$ between 200 to 250\,keV, and is $9.4~\sigma$ if the 150--200\,keV band is searched. This is the first detection of LFQPO above 200\,keV in X-ray binary systems known to us. The LFQPO frequency gradually evolved from 0.02\,Hz to 0.65\,Hz with decreasing fractional rms as the source energy spectrum becomes softer (Extended Data Fig. 1), similar to that observed in other BH binaries (BHBs)\cite{Uttley2014}. Despite a small bump below 10\,keV, the LFQPO fractional rms is almost flat in different energy bands (Extended Data Fig. 2).

\begin{figure}
\centering
\includegraphics[width=0.9\textwidth]{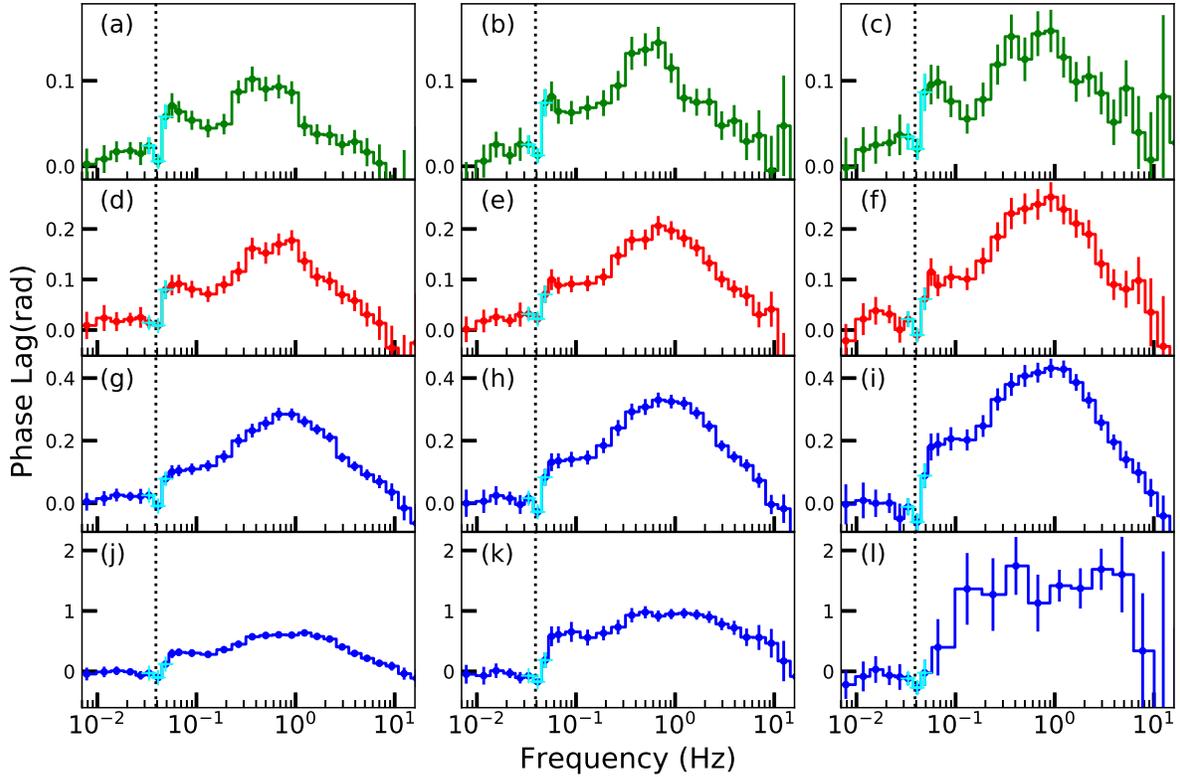}
\caption{Frequency-dependent phase-lag spectra for MAXI J1820+070 in different energy bands relative to the 1--2.6\,keV (LE) band, for a typical observation taken on MJD 58199.5--58200.9 (ObsID P0114661003): (a) 2.6--4.8\,keV, (b) 4.8--7\,keV, (c) 7--11\,keV for LE; (d) 7--11\,keV, (e) 11--23\,keV, (f) 23--35\,keV for ME; (g) 25--35\,keV, (h) 35--48\,keV, (i) 48--67\,keV, (j) 67--100\,keV, (k) 100--150\,keV, (l) 150--200\,keV for HE. In order to compare the results from different telescopes, we use some overlapping energy bands, i.e. (c) for LE and (d) for ME, (f) for ME and (g) for HE. The frequency-dependent phase-lag spectra confirm a good consistency between different telescopes. The vertical dashed lines indicate the LFQPO frequency. In all cases, we observe a narrow dip-like feature (cyan points) at the LFQPO frequency.}
\label{figure2}
\end{figure}

The properties of high-energy LFQPO are crucial to probe the nature of the accretion flow in the central regions. In particular, the LFQPO phase lag can provide insights into the geometry and the radiation processes of the accretion flow, and has important implications for the LFQPO origin. Therefore, in the following sections we will concentrate on the behaviors of the LFQPO phase lag. We produce the frequency-dependent phase-lag spectra (i.e. lag-frequency spectra) for different energy bands, with reference to the 1--2.6\,keV band (see {\bf Methods}). The lag-frequency spectra are similar between different energy bands (Figure \ref{figure2}), and also exhibit similar dependencies on the frequency among different observations (Extended Data Fig. 3(a)). The only apparent difference between the observations is that the phase lag shows an increasing trend above $\sim 10$\,keV as its energy spectrum becomes softer. A narrow dip-like feature at the LFQPO frequency (Figure \ref{figure2} and Extended Data Fig. 3(a)) is always observed and its depth increases with energy, reaching to a depth of 0.2\,rad ($\sim$0.9\,s) in the 150--200\,keV band (see Method). Such a dip feature has been reported in other BHBs like GRS 1915+105\cite{Pahari2013,Yadav2016} and GX 339-4\cite{Zhang2017} below 20\,keV, but with a much smaller time scale (less than a few milliseconds).

\begin{figure}
\centering
\includegraphics[width=0.6\textwidth]{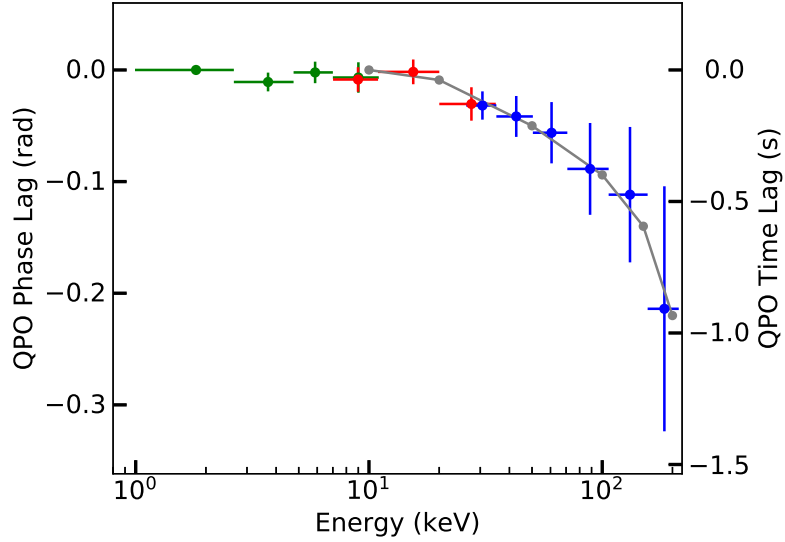}
\caption{The evolution of the LFQPO phase lag as a function of photon energy for ObsID P0114661003. In the jet precession model, as the jet is curved, different parts of the jet have different $\varphi_{\rm flow}$ (see Figure \ref{figure4}), which causes the phase lags between different energies. The curvature of the jet ($\Delta \varphi_{\rm flow}$) is tuned to match the observed phase lags (see the grey line).}
\label{figure3}
\end{figure}

The narrow dip at the LFQPO frequency is a common characteristic in the lag-frequency spectra of MAXI J1820+070. The common presence of such a narrow feature at the LFQPO peak suggests that the feature (i.e. the dip) is intrinsically associated with the LFQPO\cite{Casella2004}, and thus the production of the dip may be exclusively connected to the physical processes that produce the LFQPO. However, more observational evidence and theoretical understanding are required to robustly establish the relationship between the dip and the LFQPO. Other high energy processes, such as the inward propagation of fluctuations\cite{Arevalo2006} in the accretion flow, may contribute the broadband noise in the power density spectra (PDS) and the underlying phase-lag continuum spectra\cite{Ingram2013}. Based on the dip-LFQPO relationship assumed above, the dip feature at the LFQPO frequency can be considered as the ``intrinsic'' LFQPO phase lag, which may reveal the dominant mechanism for the production of the LFQPO. The phase lag directly measured at the LFQPO frequency from the lag-frequency spectra is referred to as the ``original'' phase lag in this paper. The average value of data points below the LFQPO frequency in the lag-frequency spectra is considered as the phase-lag continuum. Since the phase lag of the LFQPO is independent from that of the noise\cite{Ingram2019}, we can subtract the phase-lag continuum from the ``original'' phase lag, and obtain the ``intrinsic'' phase lag at the LFQPO frequency and its evolution as a function of photon energy (see {\bf Methods}), as shown in Figure \ref{figure3}, Extended Data Fig. 3 and Supplementary Table 1. The absolute value of the ``intrinsic" phase lag is the dip depth. Below, the ``intrinsic'' phase lag is referred to as the phase lag for short. The phase lag remains more or less constant at zero below 30\,keV, while above 30\,keV, the lag decreases with energy to significant soft lags and reaches a value of 0.9\,s in the 150--200\,keV band. The energy-dependent behavior at low energies ($\lesssim$30\,keV) is similar to that observed in other BHBs with RXTE\cite{Eijnden2017}. The typical time lags of these BHBs with similar LFQPO frequencies are around 0.01\,s, such as Swift J1753.5-0127\cite{Eijnden2017}, XTE J1550-564\cite{Eijnden2017}, GX 339-4\cite{Zhang2017} and radio-quiet $\rm \chi$ state of GRS 1915+105\cite{Pahari2013,Yadav2016}. Please note that these lags are not corrected for the contribution from the lag continuum, but the contribution of the lag continuum below $\sim$10\,keV for MAXI J1820+070 is negligible (see Figure \ref{figure2}, Extended Data Fig. 3(b) and (c)), we thus can compare the results directly. The time lag at high energies ($\gtrsim$30\,keV), which are reported for the first time in MAXI J1820+070, is much larger (e.g., $\sim$0.9\,s in 150--200\,keV). The large soft lag at much higher energies discovered by \emph{Insight}-HXMT observations may provide important observational diagnostics on the physical mechanisms of LFQPOs.

The main LFQPO properties of MAXI J1820+070 are summarized as follows. (1) The LFQPOs are detected at photon energies above 200\,keV; the LFQPO frequency and fractional rms are nearly constant in different energy bands; (2) the amplitude of the LFQPO soft phase lags significantly increases with energy above 30\,keV; and (3) the maximum soft lag is 0.9\,s, much larger than the time scale found previously. As discussed above, the current models proposed to explain the origin of LFQPOs are mainly based on two different mechanisms (instabilities and geometrical effects). For the latter case, precession can make the accretion flow oscillate and lead to LFQPOs.

Using these observational facts (the LFQPO's high energy, soft lag and the maximum value, and energy-related behaviors for frequency, fractional rms and phase lag), we can review and compare the existing models: instabilities in the disk\cite{Tagger1999} or in the corona\cite{Cabanac2010}; relativistic precession of the hot inner flow\cite{Ingram2009,Veledina2013,Ingram2016,Ingram2017} or the jet base\cite{Stevens2016,deRuiter2019}. If the instabilities in the accretion disk are responsible for the LFQPO\cite{Tagger1999}, it is difficult to interpret the high-energy LFQPO observed in MAXI J1820+070, especially for that with an energy above 200\,keV, as the optically thick accretion disk generally emits soft photons (which typically peak at $\lesssim 3$\,keV). We also consider the scenario that the LFQPO arises from the fluctuations in the corona\cite{Cabanac2010}. Based on a simple picture that the time lag is due to the travel-time lag, a lag of 0.9\,s would correspond to a size of $\sim 2.7 \times 10^{5} $\,km, $\sim 10^{4}$\,$r_{\rm g}$ for a 7--10\,$M_{\odot}$ BH\cite{Torres2019,Atri2020,Torres2020}. It seems that the emitting region size is unphysical so that the Comptonization of soft photons cannot explain the lag time scales. We should note that the reverberation lag is not a simple travel-time lag and the size inferred here is an upper limit. However, as the hard photons are produced by inverse Comptonization, this scenario would result in hard lags, i.e. high-energy photons lag behind low-energy ones, in direct conflict with the observations. Also, the LFQPO rms is suggested to increase with energy in this model\cite{Cabanac2010}, inconsistent with the observations. Moreover, more and more observational results in BH family, such as the inclination dependence for the LFQPO amplitude\cite{Motta2015} and the phase lag\cite{Eijnden2017}, and the modulations of the iron line energy\cite{Ingram2016} and the reflection fraction\cite{Ingram2017} in the phase-resolved spectra of H1743-322, indicate that LFQPOs may be a geometrical effect. Our findings in MAXI J1820+070 also support the geometrical effect.

The Lense-Thirring precession of the hot inner flow\cite{Ingram2009,Veledina2013,Ingram2016,Ingram2017} and the jet base\cite{Stevens2016,deRuiter2019} are two models classified as geometrical effect. In the former case, the inner flow precesses almost at the disk plane, while in the latter case, the jet precesses above the disk plane. Although the two models are two different explanations for LFQPO, both cases make the flux vary quasi-periodically due to the relativistic precession, and the inner flow and the jet can coexist in the truncated disk geometry\cite{Ingram2009,Ingram2019}. For many BHs, the Lense-Thirring model of the inner flow can well explain the observational characteristics. However, taking into account the behaviors in the HID and the HRD (see the Methods section), MAXI J1820+070 may be a peculiar source. Its behaviors in the hard state are somewhat incompatible with the picture that the Lense-Thirring precession of the inner flow dominates the variability: based on the modelling of the reflection component\cite{Buisson2019} and the appearance of broad unchanging iron line\cite{Kara2019}, the inner disk radius may be small and stable; the LFQPO frequency increases with time during the observations, with the constant inner disk radius; the LFQPO frequency is nearly constant at different energies; the emitting scale height of the hard component is much larger\cite{Buisson2019} than that in the Lense-Thirring model of the inner flow\cite{Ingram2017}. We also note that there are large uncertainties to use the reflection model to measure the inner disk radius and the results of Kara et al.\cite{Kara2019} are from a stable hard state (the disk is thus stable), thus the evidence for a constant inner disk radius is not unambiguous. Nevertheless, compared to the scenario that the Lense-Thirring precession of the jet dominates the production of the LFQPO (see below), the Lense-Thirring precession of the inner flow still faces considerable challenges, even if not excluded completely.

We then consider the scenario based on the Lense-Thirring precession of a small-scale jet. First, let us discuss the formation and the radiation processes of a small-scale jet. Among many jet models, the small-scale magnetic flux tube model proposed by Yuan et al.\cite{Yuan2019a,Yuan2019b} is applied, since it is rather generic and can explain the small hard X-ray-emitting region\cite{Kara2019,Buisson2019} of MAXI J1820+070. In their model, magnetic flux tubes in BH systems may arise due to magnetorotational instability or magnetic buoyancy. The flux tubes can form a closed zone with a size of a few gravitational radii, as a consequence of the balance between the pressure of the twisted field induced by BH spin and the confinement pressure from the external field. With the release of the magnetic energy, charged particles attached to the closed zone can be accelerated, and then, soft photons from the accretion disk can be inverse Compton scattered by the energetic electrons into X-rays. The magnetic energy dissipated in this process will be compensated by the rotational energy of the BH. This is the formation and radiation mechanisms for the small-scale jet around the BH.

Now, based on the small-scale jet precession, we will explore the physical mechanisms to produce the high-energy LFQPO and its soft lag, and explain why the LFQPO's frequency and rms are independent with energy. The jet will undergo precession if the accretion flow and thus the jet are misaligned with the BH spin axis\cite{Liska2018}, as a result of relativistic frame-dragging (i.e. Lense-Thirring precession\cite{Lense1918}). The Comptonization in the jet base would contribute high-energy photons. With the increase of distance from the BH along the jet, the maximum electron energy becomes smaller due to weakening of the magnetic field and the Compton cooling along the jet. So, the low-energy photons can originate from the top of the jet. When the jet precesses around the spin axis, the observed flux is modulated as the results of Doppler effects and solid angle effects, which generates the LFQPO. In this process, LFQPOs at different energies would be produced from different parts of the jet, i.e., high-energy LFQPO from the jet base and low-energy LFQPO from the jet top. Considering that the jet twists around the spin axis, a soft lag would be present when the jet base is observed first. Moreover, different parts (i.e. different photon energy) of the jet precess around the spin axis with the same frequency, thus such a precession can make constant LFQPO frequency in different energy bands. According to Doppler beaming, the variability amplitude in observed flux depends on the jet speed, and the jet speed does not change at different parts of the small-scale jet, in turn leading to the constant LFQPO fractional rms in different energy bands. So, for MAXI J1820+070, the LFQPO's high energy, soft lags, and energy independence for frequency and rms can be naturally interpreted in the framework of jet precession.

\begin{figure}
\centering
\includegraphics[width=0.95\textwidth]{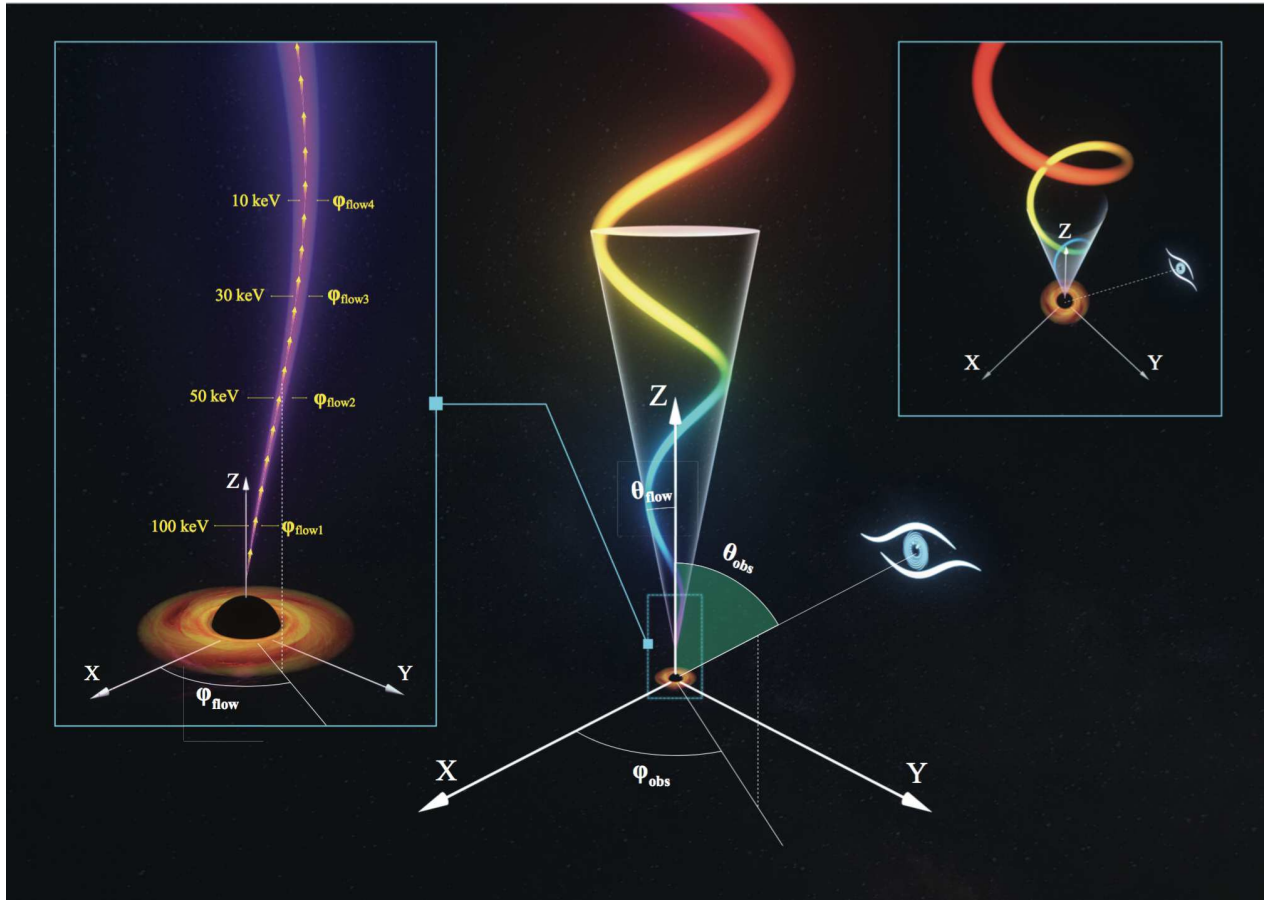}
\caption{
Schematic of the jet precession model. The black dot indicates the location of the BH, and the Z-axis follows the direction of the spin. For simplicity, we assume that the small-scale jet, marked by the colour ribbon, twists on a circular conical surface, therefore, the projected angle between the jet and the spin axis ($\theta_{\rm flow}$) is constant. $\theta_{\rm obs}$ is the inclination angle between the observer's line of sight and the BH spin axis. $\varphi_{\rm obs}$ is the projected angle of the line-of-sight on the X-Y plane. The left inset shows an enlarged version of the jet. The photon energy produced in the jet decreases with increasing jet height. $\varphi_{\rm flow}$ is the projected angle of the jet flow on the X-Y plane. Different $\varphi_{\rm flow}$ are marked along the jet in order to show that the jet is curved. By changing $\varphi_{\rm flow}$, we simulate the jet precession around the spin axis. In the right inset, the BH-jet system is rotated in order to exhibit clearly the twist of the jet around the BH's spin axis.
\label{figure4}}
\end{figure}

We then further quantitatively explain some observed timing properties (mainly the LFQPO fractional rms, the energy dependence of the soft lag and its maximum value) using a model of the jet precession. As shown in Figure \ref{figure4}, the high-energy photons above 100\,keV dominate the emission from the jet base, while the low-energy photons ($\sim$10\,keV) come mainly from the jet top. The jet twists and precesses around the spin axis of the BH. Due to Doppler boosting, the apparent brightness of the jet is determined by $\emph{D}^{p}$, where $D$ is the Doppler factor and $p$ is related to the spectral index. $D$ is a function of the jet speed ($v$) and the jet projected angle on the line-of-sight ($\theta$). $\theta$ can be calculated using the jet projected angle on the spin axis ($\theta_{\rm flow}$), the jet projected angle on the X-Y plane ($\varphi_{\rm flow}$), the inclination angle ($\theta_{\rm obs}$) and the projected angle of the line-of-sight on the X-Y plane ($\varphi_{\rm obs}$) (see Methods). During the precession, $\theta_{\rm flow}$ is assumed to be a constant, and $p$, $v$, $\theta_{\rm obs}$ and $\varphi_{\rm obs}$ do not change. Thus, $\emph{D}^{p}$ can be simplified to a function of $\varphi_{\rm flow}$. By gradually changing the phase angle $\varphi_{\rm flow}$, we can simulate the jet precession and its light curves (Extended Data Fig. 4). The LFQPO fractional rms is affected by the amplitude of the light curve, so it is determined by $v$. Assuming $\theta_{\rm obs}=~63^{\circ}$\cite{Atri2020}, $\varphi_{\rm obs}$ of $30^{\circ}$, $\theta_{\rm flow}$ of $5^{\circ}$ and $p=1-3$, the LFQPO fractional rms ($\sim 10\%$) can be reproduced (see Extended Data Fig. 2) if the jet speed is $0.48-0.99~c$. The speed is consistent with the measurement determined from the radio jet\cite{Atri2020,Bright2020}, and is also similar to the speed observed in the radio jet of some other BHBs (candidates), such as MAXI J1535-571\cite{Russell2019}, XTE J1550-564\cite{Hannikainen2009} and GRO J1655-40\cite{Hjellming1995}. Moreover, the jet precession model can also explain the energy dependence of soft lag and its maximum value. Since the jet is assumed to be twisted, the phase angle $\varphi_{\rm flow}$ is different for different part of the jet (i.e., different photon energy) and a soft phase lag is expected if the jet base is observed first. Thus, the phase lag can be converted into the difference in $\varphi_{\rm flow}$ ($\Delta \varphi_{\rm flow}$). $\Delta \varphi_{\rm flow}$ increases along the jet, so that the soft lag will increase with energy. This is the explanation for energy dependence of the phase lag in our jet model. As shown in Figure \ref{figure3}, by changing $\Delta \varphi_{\rm flow}$, we can reproduce the observed phase lag in different energy bands. The maximum lag of 0.9\,s corresponds to a phase lag of 0.21\,rad, which means that $\Delta \varphi_{\rm flow}$ between the jet base and the jet top is about 12$^{\circ}$.

In addition, the jet model can naturally explain the timing and the spectral evolutions through changing the jet size and velocity. For example, as shown in Phase B of Extended Data Fig. 1, the LFQPO frequency increases when the hardness ratio decreases. If the jet contracts by decreasing its height during this period, the precession frequency would increase as predicted by the Lense-Thirring model\cite{Fragile2007,Ingram2009}, and the spectra would soften because more hard photons from the jet are reflected by or reprocessed in the accretion disk. This ``contraction" scenario has been found with the timing analysis of \emph{NICER} data\cite{Kara2019} and the reflection modeling of \emph{NuSTAR} data\cite{Buisson2019}. In Phase C, the LFQPO fractional rms decreases. As discussed above, the LFQPO rms is determined by $v$, $p$, $\theta_{\rm flow}$, $\theta_{\rm obs}$ and $\varphi_{\rm obs}$. With a simple assumption that $p$, $\theta_{\rm flow}$, $\theta_{\rm obs}$ and $\varphi_{\rm obs}$ do not change in these observations, the evolution of the LFQPO rms can be explained as the result of the change in the jet speed. However, the physical mechanism responsible for changing either/both the size or/and the velocity of the jet needs to be further explored.

Our jet model is also in agreement with the general picture of the hard X-ray-emitting region. Based on the reverberation time lag and the modeling of the reflection component, the region is found to be located at a few gravitational radii above the BH with a lamppost geometry\cite{Kara2019,Buisson2019}. The jet size, suggested by the magnetic flux tube model\cite{Yuan2019a,Yuan2019b}, is usually a few gravitational radii, consistent with the height above. Moreover, since the projected area of the jet top on the accretion disk is larger than that of the jet base, most reflection component and the iron line will be produced by illumination from the jet top. In such a way, `lamppost' above the accretion disk is a good approximation of the jet shown in Figure \ref{figure4}.

The small-scale jet can be accelerated and collimated into a relativistic, large-scale jet by the magnetic field\cite{Blandford1977, Blandford1982}. The large-scale jet will power broadband synchrotron radiation from optical to radio. Optical radiation is generated at the start of the acceleration process, while radio radiation is generated in a region far away from the central object. We should expect that there are some multi-wavelength observational evidences in support of this picture. Radio emission is detected throughout the hard state, indicating the presence of a large-scale relativistic jet\cite{Trushkin2018,Atri2020}. The optical and near-infrared (IR) fluxes, inferred from the multi-wavelength SED and the fast optical/IR variability\cite{Shidatsu2018,Townsend2018}, are likely to be jet-dominated, and the optical emission region may have a maximum height of $\sim 5000$\,$r_{\rm g}$ above the jet base\cite{Paice2019}. More importantly, the LFQPO signal is found in fast optical observations with a centroid frequency similar to that seen in the X-ray band\cite{Yu2018a,Yu2018b}. The optical-emitting region in the jet is very close to that for X-ray, so the X-ray and optical flux may share similar variations. Thus, the jet precession can also produce the LFQPO observed in the optical band.

\section*{Summary and conclusions}

In this work, we reported on the discovery of low-frequency Quasi-Periodic Ossifications (LFQPOs) above 200\,keV with high significance in the X-ray hard state of MAXI J1820+070. This is the highest energy LFQPO detected in a black hole binary (BHB) so far. Below 30\,keV, the phase lag of the LFQPO remains constant around zero, which is similar to that observed in other BHBs. Above 30 keV, however, the lag becomes a soft lag (i.e. when high-energy photons proceed the lower-energy photons), and its amplitude significantly increases with energy and reaches the maximum value of $\sim$0.9\,s in the 150--200\,keV band. In addition, the LFQPO in different energy bands has a similar centroid frequency and fractional rms. The high detection energy (above 200\,keV), the soft lag, the large lag of $\sim$0.9\,s in 150--200\,keV, and the energy-related behaviors of frequency, fractional rms and phase lag, for the LFQPO, are incompatible with most existing models proposed to explain the origin of LFQPOs, but can be naturally explained as the result of a small-scale jet precession. Moreover, the jet model can well explain the timing and spectral evolutions of the source, and is consistent with the physical picture of the hard X-ray-emitting region, and is also supported by multi-wavelength results.

Our results reveal the relationship between the jet precession and the LFQPO in the high-energy band ($>30$\,keV), and open a new window to understand the physical processes that happen in jets launched from accreting black holes. These results also highlight the unprecedented capability of \emph{Insight}-HXMT in the hard X-ray bands, especially for nearby bright sources. For instance, the average count rate above 30 keV, for the outburst peak of MAXI J1820+070, is $>2000$~count~s$^{-1}$. Even above 100 keV, there are still enough photons to do detailed timing analysis (see Supplementary Table 2). Thus, we expect that \emph{Insight}-HXMT will provide new insight into our understanding of the X-ray variability in bright BHBs.



\section*{Methods}

\subsection*{Data reduction for {\it Insight}-HXMT}
{\it Insight}-HXMT\cite{Zhang2014,Zhang2020} is the first X-ray astronomy satellite of China, and was launched on June 15, 2017.
It operates in a low earth orbit of 550\,km with an inclination of 43$^\circ$. {\it Insight}-HXMT carries three main payloads onboard: the High Energy X-ray telescope (HE, 20--250\,keV, 5100\,cm$^{2}$), the Medium Energy X-ray telescope (ME, 5-30\,keV, 952\,cm$^{2}$), and the Low Energy X-ray telescope (LE, 1--15\,keV, 384\,cm$^{2}$). The time resolutions are 1\,ms for LE, 280\,$\rm \mu$s for ME and 25\,$\rm \mu$s for HE. HE consists of 18 cylindrical NaI(Tl)/CsI(Na) phoswich detectors, each with a diameter of 190 mm and a thickness of 3.5 mm and 40 mm for NaI and CsI respectively. HE normally operates in the regular mode, in which NaI works in the energy range of about 20--250 keV while CsI in about 70--800 keV. Thanks to the hard spectrum of MAXI J1820+070 and the large effective area of CsI, we also detect signal in CsI detectors (see Extended Data Fig. 5).

We use the \emph{Insight}-HXMT Data Analysis software (HXMTDAS) v2.0 to analyze the observations. The data are filtered with the following criteria: (1) pointing offset angle $< 0.05^{\circ}$; (2) elevation angle $> 6^{\circ}$; (3) the value of the geomagnetic cutoff rigidity $> 6$. The backgrounds for HE, ME and LE are estimated according to the linear correlation between the detectors with small field of view (FoV) and blind detectors, and the coefficient is the number of nonblind detectors to that of blind detectors. This method is tested by the {\it Insight}-HXMT background team using blank sky observations and is adopted in the timing analysis of MAXI J1535-571\cite{Huang2018}. For typical exposures of MAXI J1820+070, the background systematic errors of LE (1--10\,keV, 2\,ks), ME (8.9--44\,keV, 2\,ks) and HE (26--100\,keV, 4\,ks) are $3.2$\%\cite{Liao2020a}, $\sim 3$\%\cite{Guo2020} and $2.0$\%\cite{Liao2020b}, respectively. Here we need to point out that the background systematic error of ME is not directly measured for an exposure of 2\,ks. It is estimated by extrapolating the relationship between background systematic errors and exposure times reported in Guo et al.\cite{Guo2020}

\subsection*{State transition}

A typical BHB usually undergoes several different spectral states during an outburst (e.g., Homan \& Belloni\cite{Homan2005}; Remillard \& McClintock\cite{Remillard2006}; Done et al.\cite{Done2007}; Belloni \& Motta\cite{Belloni2016}, for reviews). It is in the hard state at the beginning of the outburst. The state is characterized by high fractional rms variability typically around 30--40\% and a hard spectrum dominated by a power-law component. The source then goes through an intermediate state and further evolves into the soft state. In this period, the spectra continue to soften, and the total fractional rms decrease from $\sim$30--40\% to a very low level around or less than 1--2\%. The soft state can be last for months. Then the source leaves the soft state and passes through an intermediate state, with a luminosity lower than that of the early intermediate state, and turns back to the hard state. The spectra continue to harden, and the total fractional rms increase to the same level of the previous hard state. Therefore, the source typically follows an counterclockwise q-shaped track in the hardness-intensity diagram (HID) during the outburst, and no hysteresis pattern is expected in the hardness-rms diagram (HRD). Moreover, a steady jet may be detected in the hard state, and LFQPOs are usually present in the hard and intermediate states and occasionally present in the soft state.

The HID and the HRD of MAXI J1820+070 are shown in Extended Data Fig. 6. The total fractional rms (1--10\,keV, LE) is computed within the frequency bands of 0.01--32\,Hz. The data points of the HID track a large q-shaped curve during the whole outburst (from phase A to F) and show a small ``q-like" track in Phase A, B and C. The HRD follows a ``$\infty$-like'' pattern. The outburst starts at Phase A and follows the small ``q-like" track in the right side of the HID for about three months, with a slight softening in spectra, corresponding to the hard state; the fractional rms remains at $\sim$30\%--$\sim$45\%. While in Phase D, the source moves to the upper left side of the HID and the fractional rms suddenly drops to $\sim$6\%, indicating that the system is in the intermediate state. Then the source moves to the left branch of the HID during Phase E, and the rms decreases to a level of $\sim$0.2\%, corresponding to the soft state. Finally, the source returns to the hard state in Phase F, and the rms increases to $\sim$40\%. The state evolution determined from the {\it Insight}-HXMT data is consistent with that from {\it MAXI} data\cite{Shidatsu2019}.

The states and transitions of MAXI J1820+070 are approximately compatible with those of other BHs, but exhibit some peculiar behaviors, such as the small ``q-like" track and a hard-to-hard transition\cite{Wang2020} in the hard state, and the ``$\infty$-like'' pattern in the HRD. It looks that MAXI J1820+070 is not a typical one among BH transients, and in-depth investigations are needed in future.

\subsection*{Power density spectra}
The Good Time Interval (GTI) files for LE, ME and HE are generated using {\tt legtigen}, {\tt megtigen} and {\tt hegtigen} tools of HXMTDAS, respectively. Applying these GTI files, the LE, ME and HE event files are respectively filtered using {\tt lescreen}, {\tt mescreen} and {\tt hescreen}. We produce PDS from 256\,s data intervals with a time resolution of 1/128s for each observation. The average power spectra are computed in the energy band 1--10\,keV for LE, 10--30\,keV for ME and 35--250\,keV for HE, using {\tt powspec} of HEASOFT. No high-frequency QPOs appear in this source, we thus only focus on the behavior in low frequency, i.e.,$<$10\,Hz. The PDS are subjected to fractional rms-squared normalization\cite{Belloni1990} after subtracting the Poisson noise, and are re-binned in increments of 3\% of the frequency. No dead time correction is applied in the PDS, since dead time only has an impact on white noise and should not be an issue in LFQPO analysis\cite{Huang2018}. Moreover, in \emph{Insight}-HXMT, dead time ($\tau_d$) is around 20\, $\mu$s for HE and LE and 250 $\mu$s for ME, thus the frequency range commonly analyzed in BHs is well below $1/\tau_d$. Given the small dead time of HE and LE, our results will not be affected by the dead time. For ME, the deficit of the Leahy normalized power density of white noise, corresponding to a power of 2, is $\sim1.9$\% and less than the background uncertainty ($\sim3$\%), thus the dead time effect can be reliably ignored.

The average PDS between 0.002\,Hz and 8\,Hz are fitted with a sum of Lorentzians\cite{Belloni2002} using the XSPEC v12.9.1 software package. The best-fit reduced $\chi^2$ is less than 1.5, with a typical value of 1.2. The PDS are very similar to those observed in the hard state of other BHs (e.g., GX 339-4), and an LFQPO is found in the PDS between MJD 58197 to MJD 58306 (from Phase A to Phase D in Extended Data Fig. 6). The LFQPO shows an evolution in the outburst, as shown in Extended Data Fig. 1. Its frequency increases from 0.02\,Hz (MJD 58194, Phase A) to 0.51\,Hz (MJD 58257, end of Phase B), and then decreases to 0.22\,Hz (MJD 58286, end of Phase C). When the source enters into the intermediate state (phase D), the frequency increases again. The coherence parameter ($Q=\nu/\Delta\nu$) is between $\sim$3 and 7. The LFQPO significance for the longest observation (P0114661004), defined as the integral of the Lorentzian used to fit the LFQPO divided by its error, are $9.4~\sigma$ for the 150--200\,keV band and $\sim 4~\sigma$ between 200 to 250\,keV. The LFQPO fractional rms is around 10\% until MJD 58257, but decreases slowly after then.

\subsection*{Energy dependence of LFQPO parameters}
In order to quantitatively study the energy-dependence of the LFQPO properties, we obtain power spectra in 13 energy bands: LE (1--2.6\,keV, 2.6--4.8\,keV, 4.8--7\,keV, 7--11\,keV), ME (7--11\,keV, 11--23\,keV, 23--35\,keV) and HE (25--35\,keV,  35--48\,keV, 48--67\,keV, 67--100\,keV, 100--150\,keV, 150--200\,keV). The background contribution is considered in the LFQPO fractional rms calculation. The LFQPO rms is calculated in the form as rms = $\sqrt{P}*(S+B)/S$, where $S$ and $B$ are source and background count rates, respectively, and $P$ is the power normalized according to Belloni et al.\cite{Belloni1990}.

The LFQPO is found in 76 \emph{Insight}-HXMT observations from MJD 58197 to MJD 58306. The results from a typical observation on MJD 58200 (ObsID P0114661003, first red line in Extended Data Fig. 1) are shown in Extended Data Fig. 2. The LFQPO frequency is almost constant and independent of photon energy. The LFQPO fractional rms remains constant above 10\,keV, while below 10\,keV, the LFQPO rms shows a complex evolution with energy, which may be due to the hybrid contributions from the jet and the accretion disk.

\subsection*{Phase lag}
Using the method of Vaughan \& Nowak\cite{Vaughan1997} and Nowak et al.\cite{Nowak1999}, the phase lags, with reference to the 1--2.6\,keV band, are computed in 12 energy bands, i.e., LE (2.6--4.8\,keV, 4.8--7\,keV, 7--11\,keV), ME (7--11\,keV, 11--23\,keV, 23--35\,keV) and HE (25--35\,keV, 35--48\,keV, 48--67\,keV, 67--100\,keV, 100--150\,keV, 150--200\,keV). The 1--2.6\,keV band is used as the reference band, because it covers channels 100--300 of the LE telescope and is dominated by the soft component. If a broad reference band (i.e. 1--10\,keV or every other energy channel in the 1--10\,keV band) is adopted, the lag-frequency spectra and the energy dependence of the phase lag are consistent with those using the 1--2.6\,keV band. Errors in the phase lags are determined following Eqs. (16) and (17) of Nowak et al.\cite{Nowak1999}. A positive lag means that the hard photons lag the soft ones. The typical frequency-dependent phase-lag spectra are shown in Figure \ref{figure2}. We find that the lag-frequency spectra between different energy bands are similar. A narrow dip-like feature always appears at the LFQPO frequency, and its depth increases with energy. Such a feature has also been reported in other BHBs, such as GRS 1915+105 and GX 339-4, but the detections are below 20\,keV and with a much smaller scale. Above the LFQPO frequency, the phase lag first increases to its maximum, and then gradually decreases with Fourier frequency. The lag-frequency spectra are compared to the results of \emph{NICER} data. As shown in Extended Data Fig. 7, our results are consistent with those obtained in the \emph{NICER} quasi-simultaneous observation (the red spectrum of Fig. 2 in Kara et al.\cite{Kara2019}).

The lag-frequency spectra from three typical observations are shown in Extended Data Fig. 3(a). We can see that the lag-frequency spectra in different observations also have similar shapes, i.e., a dip-like feature at the LFQPO frequency and a hump following the dip, but show an overall increasing trend when the source softens (from top to bottom in Extended Data Fig. 3(a)).

The real part and the imaginary part in the complex plane are averaged respectively to calculate the phase lags. Then, for a given energy band, we average the phase lags over the LFQPO frequency range $\nu \pm \rm FWHM/2$ to determine the phase lag in this energy band. Performing the same procedure in different energy bands, we obtain the evolution of the phase lags as a function of photon energy. In order to distinguish from the lags below, the phase lags are referred to as ``original'' lag. As shown in Extended Data Fig. 3(b), with the evolution of the outburst at different stages, the ``original'' phase lags exhibit three different dependencies successively: first decreasing with energy, then remaining nearly constant, and increasing with energy at last.

Upon closer inspection (see Extended Data Fig. 3(a)), we find that the narrow dip is always detected at the LFQPO frequency and the change in the phase lags is mainly due to the overall enhancement of the lag-frequency spectra during the evolution. We are interested in the ``intrinsic'' lag directly related to the LFQPOs. Since the dip is the common feature at the LFQPO frequency and may be connected to the physical mechanism to make the LFQPO, it is considered as the ``intrinsic'' lag. Other high energy processes, which contribute the broadband noise in the power density spectra and may also be responsible for the phase-lag continuum and the overall enhancement of the lag-frequency spectra, should thus be removed. In order to measure the ``intrinsic'' lag, we need to remove the phase-lag continuum. Considering that both the power density spectra and the lag-frequency spectra are flat below the LFQPO frequency, we use the average value of all data points below the LFQPO frequency as the phase-lag continuum (see purple points in Extended Data Fig. 3(a)). The ``intrinsic'' phase lag is obtained by subtracting the phase-lag continuum from the ``original'' phase lag at the LFQPO frequency, i.e. ``intrinsic'' phase lag $=$ ``original'' phase lag $-$ phase-lag continuum. The uncertainty in the ``intrinsic'' lag is estimated by propagating the errors in the ``original'' lag and in the phase-lag continuum. The absolute value of the ``intrinsic'' phase lag is the depth of the dip. As shown in Extended Data Fig. 3(c), the ``intrinsic" phase lags follow a similar evolution with energy. They remain constant at around zero below 30\,keV, and increase to soft lags (i.e. the high-energy photons arrive first, and the low-energy photons lag behind and arrive after) above 30\,keV. The significance of the ``intrinsic'' phase lags at high energies (above 30\,keV) for a typical observation (ObsID P0114661003, $\sim 61$\,ks for HE) are $\gtrsim 2 \sigma$ (see Supplementary Table 1). The significance of the ``original'' phase lags are larger than $3 \sigma$ above 60\,keV. As the uncertainties from the phase-lag continuum contribute a significant part of the errors, the significance of the ``intrinsic'' phase lags are reduced. Taking these into account, the detections of large soft lags at high energies are reliable.

\subsection*{Jet precession model}
Figure \ref{figure4} shows the geometry of the jet. The jet twists and rotates around the BH spin axis. In this process, Doppler boosting causes the modulation of the observed flux from the jet. The observed flux of the jet is determined by the Doppler factor, $\emph{D}$, which is a function of the jet velocity ($v$) and the projected angle to the line-of-sight ($\theta$). In order to calculate $\theta$, we define a coordinate system in $xyz$ directions with the $z$-axis aligned with the BH spin, and the $xy$ plane perpendicular to the $z$-axis. The position of the observer (in $xyz$ coordinates), denoted as a vector
$\mathbf{\hat{e}}_{\rm obs}$, can be described by
\begin{equation}
\mathbf{\hat{e}}_{\rm obs} = (\sin \theta_{\rm obs} \cos \varphi_{\rm obs}, \sin \theta_{\rm obs} \sin \varphi_{\rm obs}, \cos \theta_{\rm obs}),
\end{equation}
where $\theta_{\rm obs}$ is the inclination angle between the line of sight of the observer and the BH spin axis and $\varphi_{\rm obs}$ is the azimuth angle of the observer measured from the $x$-axis. The jet flow in the $xyz$ coordinates can be written as
\begin{equation}
\mathbf{\hat{e}}_{\rm flow} = (\sin \theta_{\rm flow} \cos \varphi_{\rm flow}, \sin \theta_{\rm flow} \sin \varphi_{\rm flow}, \cos \theta_{\rm flow}),
\end{equation}
where $\theta_{\rm flow}$ is the inclination angle and $\varphi_{\rm flow}$ is the azimuth angle of the jet flow measured from the $x$-axis. $\theta$ is given by a scalar product $\cos \theta$ = $\mathbf{\hat{e}}_{\rm obs} \mathbf{\hat{e}}_{\rm flow}$, which can be expanded as
\begin{equation}
\cos \theta = \sin \theta_{\rm obs} \cos \varphi_{\rm obs}\sin \theta_{\rm flow} \cos \varphi_{\rm flow}+ \sin \theta_{\rm obs} \sin \varphi_{\rm obs}\sin \theta_{\rm flow} \sin \varphi_{\rm flow}+ \cos \theta_{\rm obs}\cos \theta_{\rm flow}.
\end{equation}

Assuming a simple jet model with a homogeneous sphere, the observed luminosity is proportional to the intrinsic luminosity as
\begin{equation}
 \emph{S}_{o} = \emph{S}_{e}\emph{D}^{p}
\end{equation}
where
\begin{equation}
p = \kappa - \alpha.
\end{equation}
$\kappa$ is 2 for continuous jet and 3 for discrete ejection, and $\alpha$ is the spectral index. The relativistic boosting factor is given by
\begin{equation}\label{Dboosting}
  \emph{D} = \frac{1}{\gamma (1-\beta \cos \theta)},
  \end{equation}
  \begin{equation}\label{Dboosting}
  \gamma = \frac{1}{\sqrt{1-\beta^{2}}},
  \end{equation}
  \begin{equation}\label{Dboosting}
  \beta = \upsilon / c,
\end{equation}

The observed luminosity depends on $v$ and $\theta$ through the Doppler factor, $\emph{D}$, and the spectral index $\alpha$. For simplicity, we assume $\theta_{\rm flow}$ to be a constant, and consider that $v$, $\alpha$, $\theta_{\rm obs}$ and $\varphi_{\rm obs}$ should not change, therefore, the observed luminosity is determined by $\varphi_{\rm flow}$.

The inclination of MAXI J1820+070 inferred from the radio jet is $63\pm 3^{\circ}$\cite{Atri2020}, consistent with the presence of X-ray dips in the light curves\cite{Kajava2019} and a likely grazing eclipse of the accretion disk\cite{Torres2019,Torres2020}. Thus, $\theta_{\rm obs}$ = $63^{\circ}$ is a reasonable assumption. Moreover, based on population synthesis models and taking the natal kick during the formation of BH into account, most Galactic BHBs should have small misalignment angles ($\lesssim 10^{\circ}$)\cite{Fragos2010}. Therefore, an $\theta_{\rm flow}$ of $5^{\circ}$ is adopted. In addition, we use $\kappa=2$ for a continuous jet, and assume that $\alpha=-1$, 0 and 1, respectively, as the spectral fitting is not performed in this paper. Using $\theta_{\rm obs}$ = $63^{\circ}$, $\varphi_{\rm obs}$ = $30^{\circ}$, $\theta_{\rm flow}$ = $5^{\circ}$, we simulate the jet precession by changing $\varphi_{\rm flow}$, and test three cases with $p =$1, 2 and 3. The simulated light curves for $p=1$ at four different energies are shown in Extended Data Fig. 4. By tuning $\varphi_{\rm flow}$ and taking into account a jet velocity of $0.48~c$, $0.65~c$ and $0.99~c$ (for $p=$3, 2 and 1, respectively), our jet precession model can reproduce the observed phase lag (see Figure \ref{figure3}) and the LFQPO fractional rms ($\sim$10\%, Extended Data Fig. 2) above 10\,keV. Since the detected photon number is insufficient for performing a QPO phase resolved analysis and constraining the jet opening angle, we do not consider the opening angle in our model.

\subsection*{LFQPO detected in the CsI detectors}
The PDS from the NaI and CsI detectors are shown in Extended Data Fig. 5. Besides the NaI detectors, the LFQPO signal is also detected in the 150--250\,keV band by the CsI detectors. The count rate in the CsI detectors is 56\,cts~s$^{-1}$. The centroid frequency of the LFQPO detected by the CsI detectors is consistent with that by the NaI detectors. The results of the CsI detectors also confirm the detection of the LFQPO above 200\,keV.

\section*{Data availability}
The data that support the plots within this paper and other findings of this study are all publicly available for download at the \emph{Insight}-HXMT website (http://www.hxmt.cn/ or http://hxmt.org/).

\section*{Code availability}
The \emph{Insight}-HXMT data reduction was done using the software which is available at the \emph{Insight}-HXMT website (http://www.hxmt.cn/ or http://hxmt.org/). The model fitting of power spectra was completed with XSPEC, which is available at the HEASARC website (https://heasarc.gsfc.nasa.gov/xanadu/xspec/). The phase lag was performed with Stingray (see https://stingray.readthedocs.io/en/latest/index.html).

{}

\section*{Acknowledgements}
We appreciate the anonymous referees for useful comments and suggestion that have helped us to improve the paper. We thank Adam Ingram, Konstantinos Karpouzas and Mariano Mendez for useful suggestions. This work made use of the data from the Insight-HXMT mission, a project funded by China National Space Administration (CNSA) and the Chinese Academy of Sciences (CAS). The Insight-HXMT team gratefully acknowledges the support from the National Program on Key Research and Development Project (Grant No. 2016YFA0400800) from the Minister of Science and Technology of China (MOST) and the Strategic Priority Research Program of the Chinese Academy of Sciences (Grant No. XDB23040400). The authors thank supports from the National Natural Science Foundation of China under Grants U1838111, U1838115, U1838201, U1838202, 11473027, 11633006, 11673023, 11733009, the National Key Research and Development Program of China (grant No. 2016YFA0400704) and the Royal Society Newton Funds.

\section*{Author contributions}
X.M., L.T., S.-N.Z., L.Z., Q.-C.B., M.-Y.G., and J.-L.Q.were involved in the presented analysis. S.-N.Z., L.T., J.-L.Q., F.Y. and F.-G.X. contributed to the theoretical discussions. The manuscript was produced by L.T., S.-N.Z., X.M., L.Z., Q.-C.B., J.-L.Q., M.-Y.G., S.Z., F.-J.L., L.-M.S.and Y.-J.Y. The PI of the \emph{Insight}-HXMT mission is S.-N.Z. All other authors contributed to the development of the mission concept and/or construction and testing of Insight-HXMT.

\section*{Competing interests}
The authors declare no competing financial interests.

\section*{Additional information}
\textbf{Correspondence and requests for materials} should be addressed to S.-N.Z.and L.T.(email: zhangsn@ihep.ac.cn, taolian@ihep.ac.cn).\\
\textbf{Reprints and permissions information} is available at www.nature.com/reprints.

\setcounter{figure}{0}
\captionsetup[figure]{labelfont={bf},labelformat={default},labelsep=period,name={Extended Data Fig.}}

\begin{figure}
\centering
\includegraphics[width=0.95\textwidth]{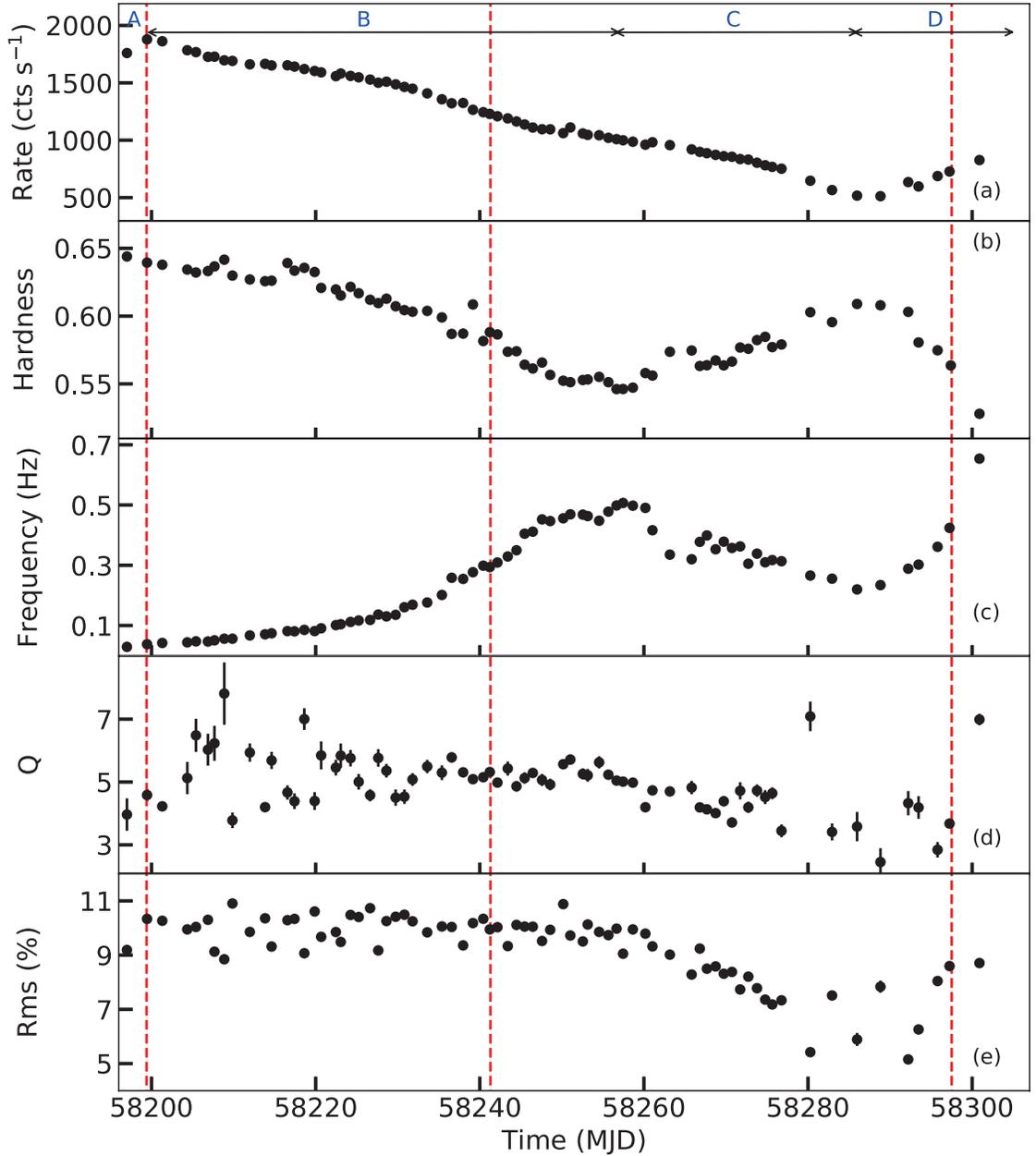}
\caption{Light curve, hardness ratio, LFQPO's frequency, Q factor and fractional rms of MAXI J1820+070 in the X-ray hard state. (a) \emph{Insight}-HXMT/HE light curve (35--200\,keV) of MAXI J1820+070 in the hard state from MJD 58190 to MJD 58301. (b) is the evolution of the hardness ratio (defined as the ratio of the net count rate in the 3.0--10.0 keV to 1.0--3.0\,keV bands). Panels (c)-(e) show the evolution of the LFQPO's frequency, Q factor and fractional rms. Phases A to D are marked in the top panel. The red dashed lines indicate the three typical observations in Extended Data Fig. 3.
}
\end{figure}

\begin{figure}
\centering
\includegraphics[width=0.95\textwidth]{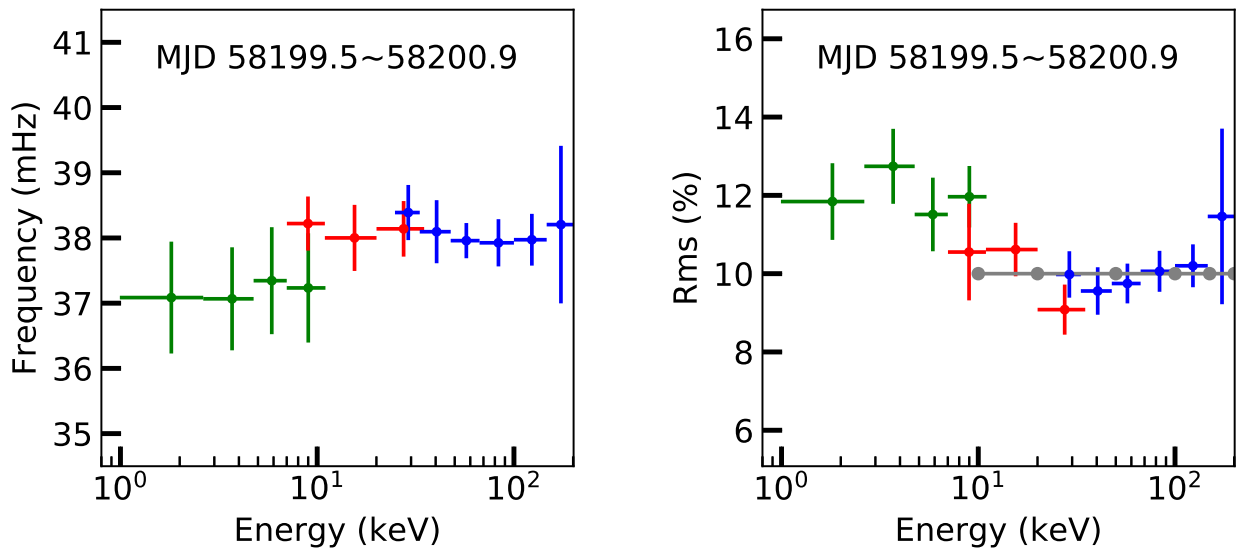}
\caption{LFQPO centroid frequency (left) and fractional rms amplitude (right) as a function of energy for a typical observation (ObsID P0114661003). The LFQPO rms is calculated in the full frequency range of the PDS. The fractional rms-squared normalization depends on the PDS, and the Lorentzian functions are used to fit the PDS. The green, red and blue points represent LE, ME and HE data, respectively. The gray points indicate the LFQPO rms values from the jet precession model with $p=1$.
}
\end{figure}

\begin{figure}
\centering
\includegraphics[width=0.95\textwidth]{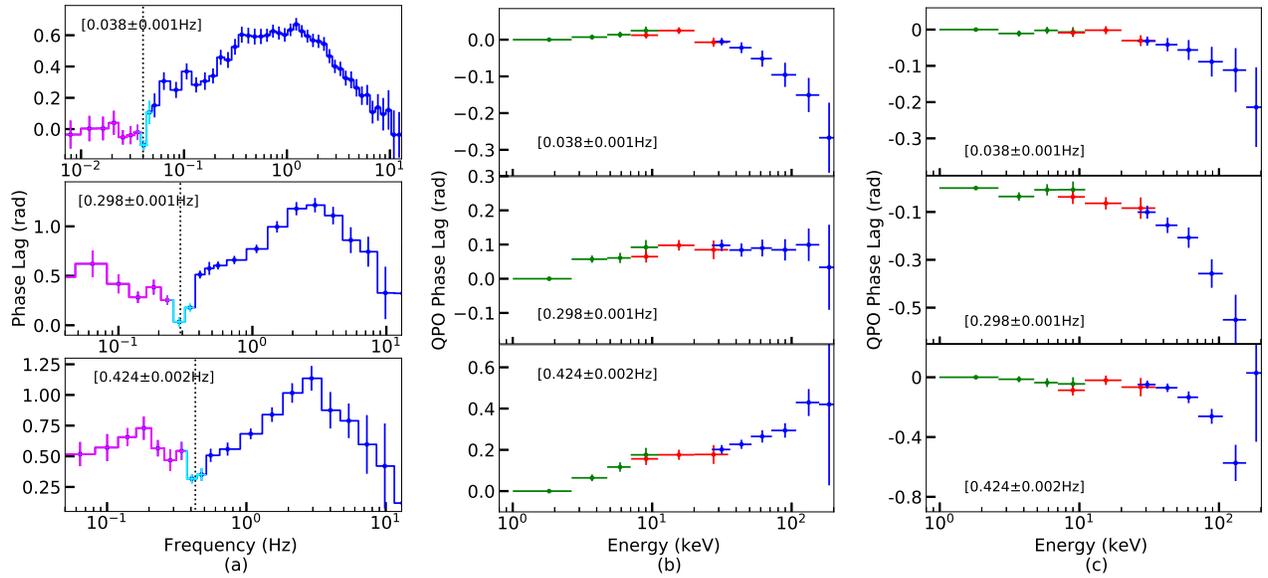}
\caption{Frequency-dependent phase-lag spectra and LFQPO phase lags for three typical observations. The observations are marked with red vertical dashed lines in Extended Data Fig. 1, and their general properties are listed in Supplementary Table 3. The spectra of the three observations continue to soften. (a) Frequency-dependent phase-lag spectra in the 67--100\,keV band. The vertical dashed lines mark the LFQPO frequency, and the cyan points show the narrow dip-like feature. (b) The ``original" LFQPO phase lags relative to the 1--2.6\,keV band. By averaging the phase lags over the LFQPO frequency range $\nu \pm \rm FWHM/2$ in different energy bands, we obtain the ``original" LFQPO phase lags as a function of photon energy. (c) The ``intrinsic" LFQPO phase lags, which are determined using the ``original" LFQPO phase lags minus the phase-lag continuum. The average value of data points below the LFQPO frequency (purple points in panel (a)) is used as the phase-lag continuum.
}
\end{figure}

\begin{figure}
\centering
\includegraphics[width=0.95\textwidth]{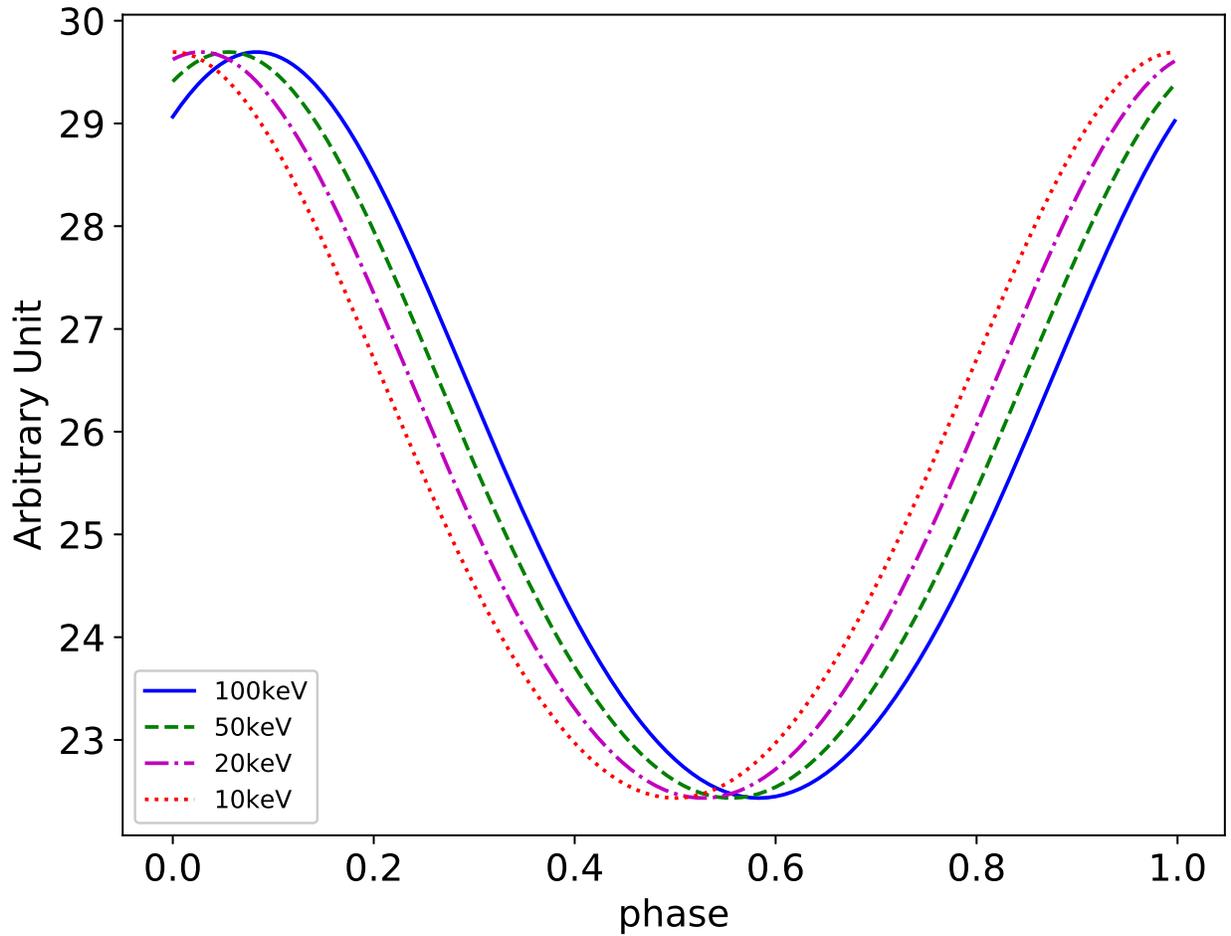}
\caption{Simulated light curves at four energies for $p=1$: 10\,keV, 50\,keV, 100\,keV, 200\,keV. From Eqs. (3) and (4), we simulate the light curves by changing the phase angle $\varphi_{\rm flow}$, assuming $\theta_{\rm flow}$, $v$, $\alpha$, $\theta_{\rm obs}$ and $\varphi_{\rm obs}$ to be constant. Using the light curves, we can calculate the jet speed based on the observed LFQPO fractional rms (see Extended Data Fig. 8).
}
\end{figure}

\begin{figure}
\centering
\includegraphics[width=0.95\textwidth]{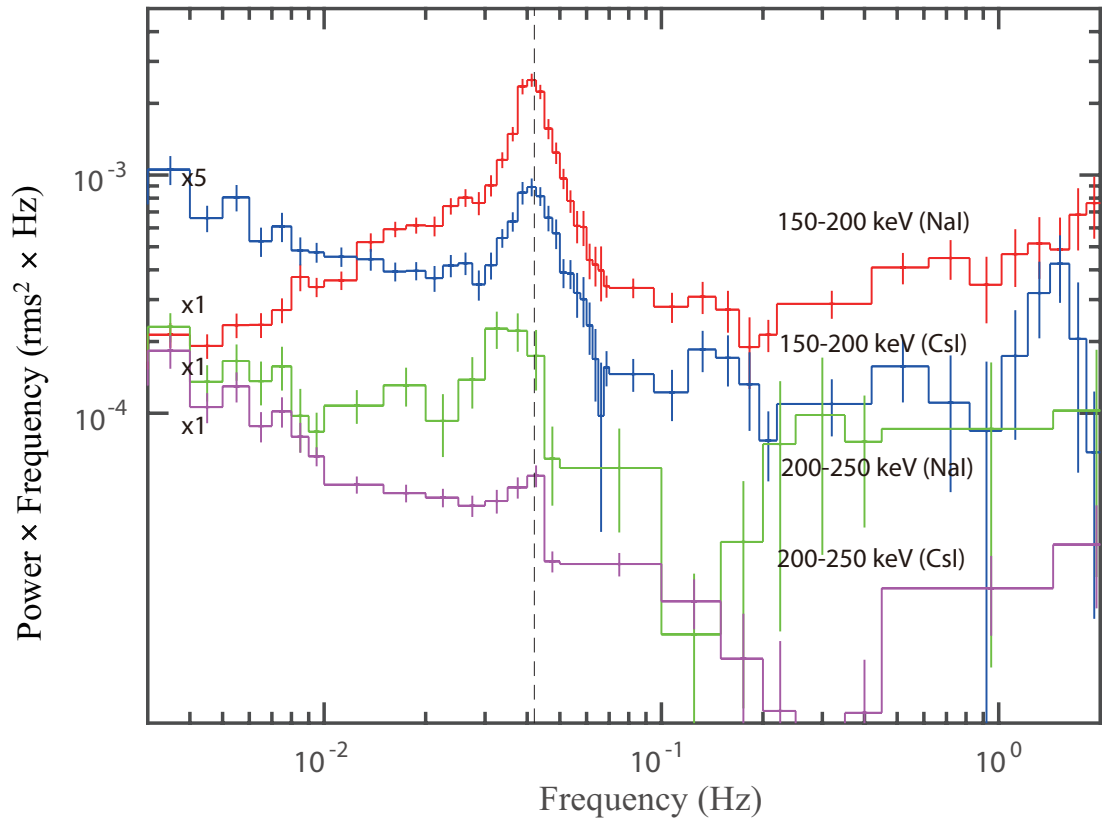}
\caption{The PDS of MAXI J1820+070 in the 150--200 and 200--250\,keV bands with different detectors: NaI and CsI, for ObsID P0114661004. The power is multiplied by a different factor for plotting clarity.
}
\end{figure}

\begin{figure}
\centering
\includegraphics[width=0.95\textwidth]{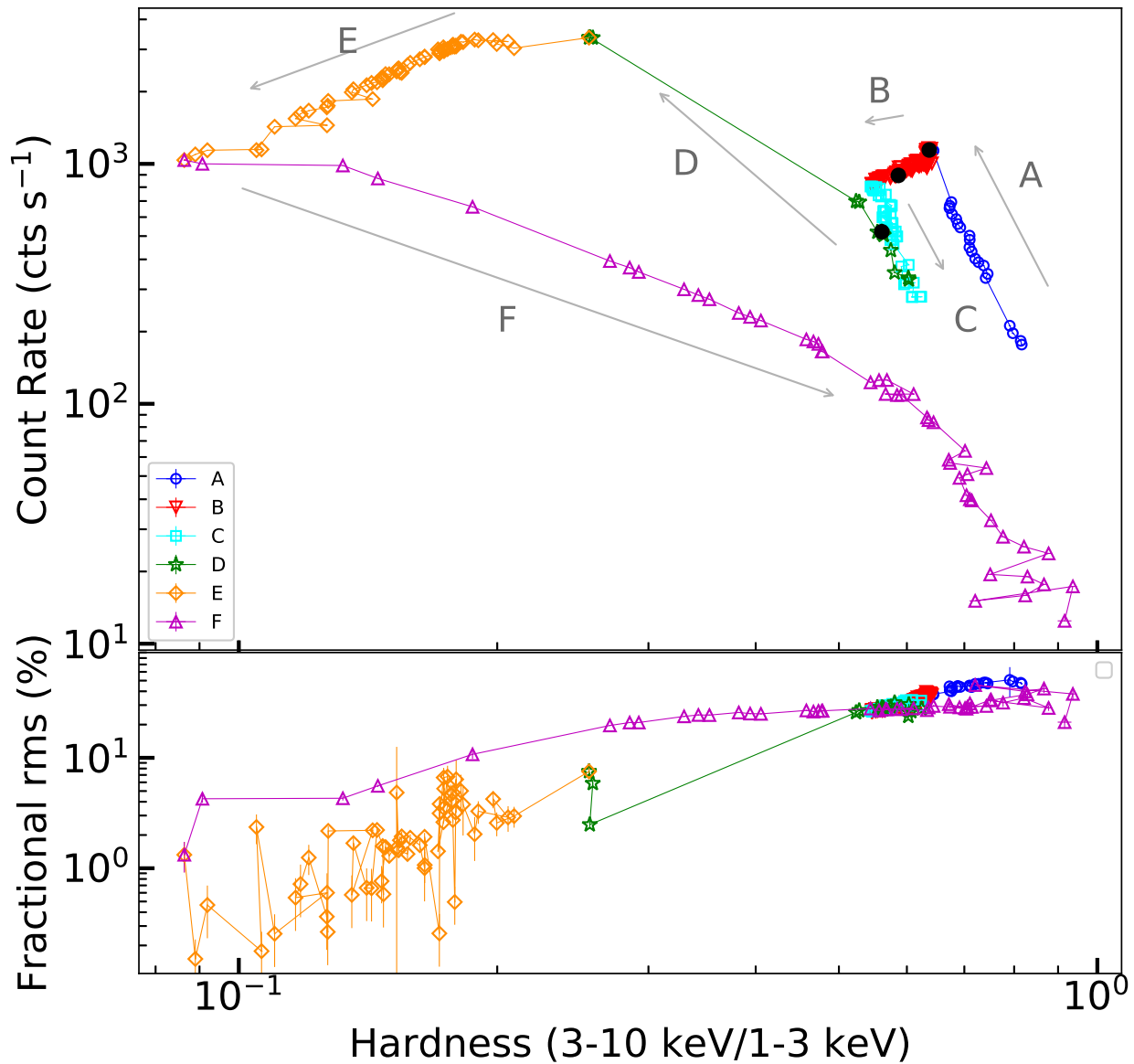}
\caption{\emph{Insight}-HXMT hardness-intensity diagram (HID) (upper) and hardness-rms diagram (HRD) (lower) of MAXI J1820+070. Each point represents a single \emph{Insight}-HXMT exposure. Data points during the six phases (A to F) are shown in different symbols. The intensity is the LE count rate in the 1.0--10.0\,keV band. The hardness ratio is defined as the ratio of the net count rate in the 3.0--10.0 keV to 1.0--3.0\,keV bands. The total fractional rms is calculated in the 0.01--32 Hz frequency range. Arrows show the evolutionary track of the outburst. The black points show the three typical observations used to calculate the LFQPO phase lag (Extended Data Fig. 3).
}
\end{figure}

\begin{figure}
\centering
\includegraphics[width=0.95\textwidth]{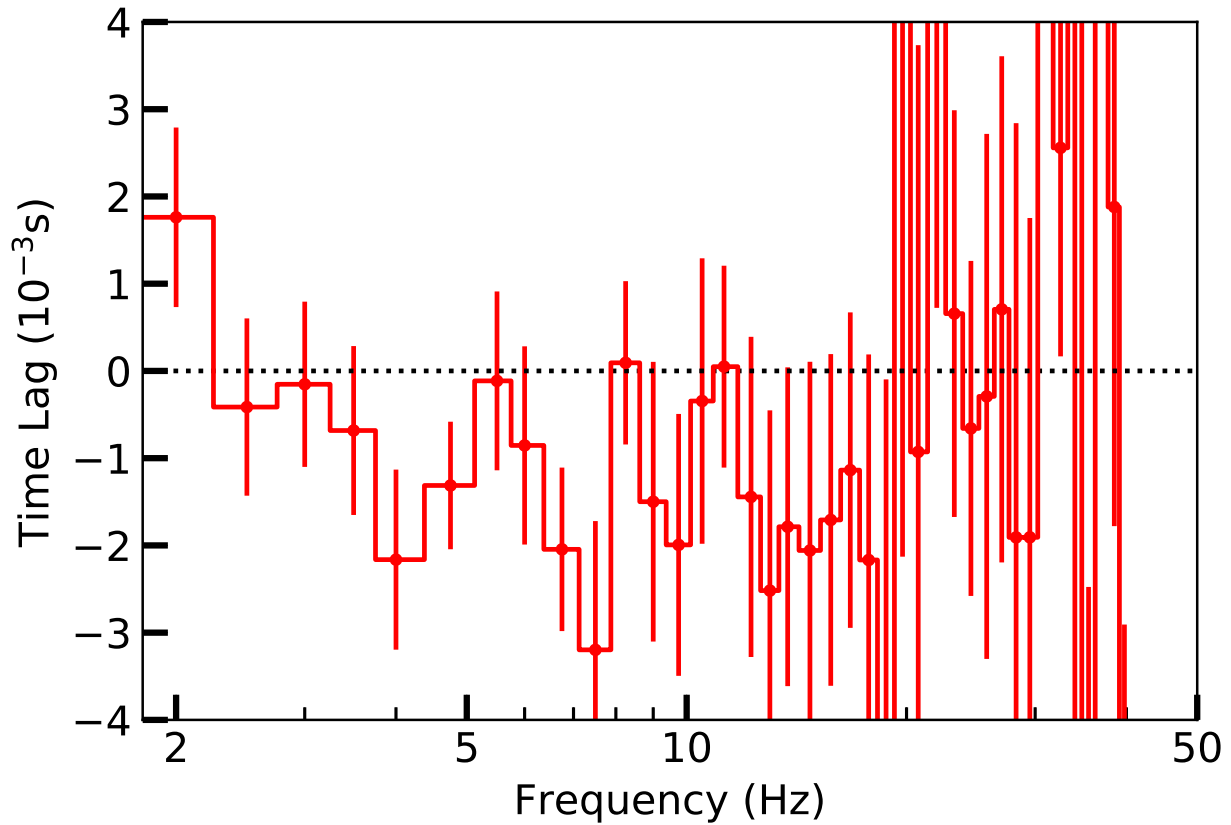}
\caption{One frequency-dependent time-lag spectrum of \emph{Insight}-HXMT. To test the consistency between \emph{Insight}-HXMT and \emph{NICER}, we make the Frequency-dependent time-lag spectrum between 0.7--1 keV and 1--10 keV for ObsID P0114661003 (MJD 58199.5--58200.9). The spectrum is consistent with the result from a quasi-simultaneous \emph{NICER} observation taken within one day (ObsID 1200120106, see the red spectrum of Fig. 2 in Kara et al. \cite{Kara2019}).
}
\end{figure}

\begin{figure}
\centering
\includegraphics[width=0.95\textwidth]{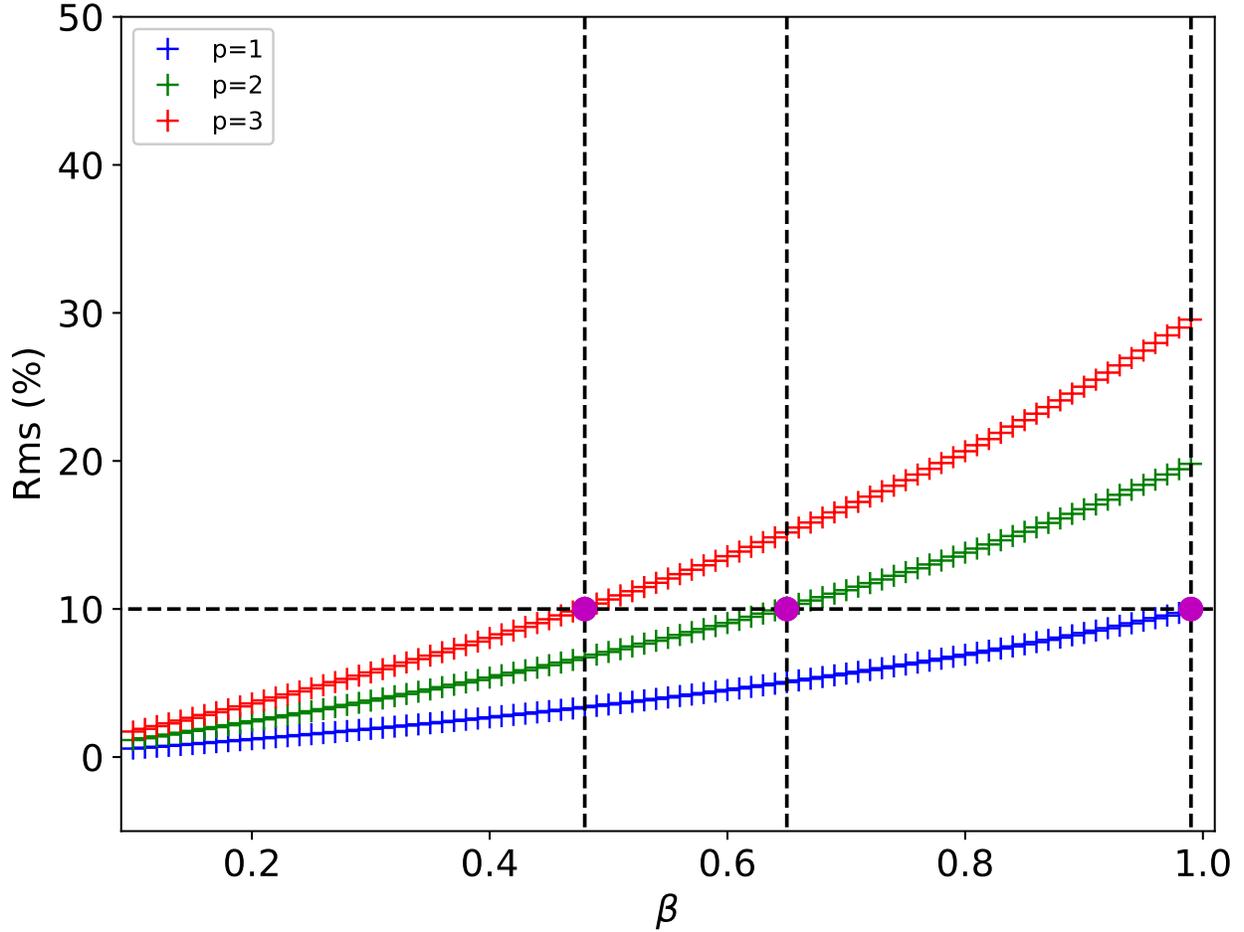}
\caption{The relations between $\beta$ and the LFQPO rms for $p=$1, 2 and 3, respectively. As discussed above, the rms is determined by the amplitude of the light curves, which is a function of $v$. We change $v$ in the range of (0.01-0.99)$~c$ with a step length of 0.01$~c$, and simulate light curve for each $v$. The rms are calculated from these light curves, so we can obtain the relation between $v$ and the rms. Using the observed rms ($\sim$10\%), the jet speed can be inferred.
}
\end{figure}
\clearpage

\captionsetup[table]{labelfont={bf},labelformat={default},labelsep=period,name={Supplementary Table}}

\begin{table}
\footnotesize
\caption{LFQPO phase lags and time lags of different energy bands for the observation taken on MJD 58199.5--58200.9 (ObsID P0114661003). }
\label{table:1}
\medskip
\begin{center}
\begin{tabular}{c c c c}
\hline \hline
Telescope & Energy band & Phase lag & Time lag \\
          & (keV)       & (rad) & (s) \\
\hline
   & 2.6--4.8 & $-0.011 \pm 0.008$ & $-0.05 \pm 0.04$ \\
LE & 4.8--7   & $-0.002 \pm 0.010$ & $-0.01 \pm 0.04$ \\
   & 7--11    & $-0.007 \pm 0.014$  & $-0.03 \pm 0.06$ \\
\hline
   & 7--11 & $-0.009 \pm 0.011$ & $-0.04 \pm 0.05$ \\
ME & 11-23 & $-0.002 \pm 0.011$ & $-0.01 \pm 0.05$ \\
   & 23-35 & $-0.030 \pm 0.015 $ & $-0.13 \pm 0.06$ \\
\hline
   & 25-35   & $-0.032 \pm 0.013$ & $-0.14 \pm 0.05$ \\
   & 35-48   & $-0.04 \pm 0.02$ & $-0.18 \pm 0.08$ \\
HE & 48-67   & $-0.06 \pm 0.03$ & $-0.24 \pm 0.12$ \\
   & 67-100  & $-0.09 \pm 0.04$ & $-0.4 \pm 0.2$ \\
   & 100-150 & $-0.11 \pm 0.06$ & $-0.5 \pm 0.3$ \\
   & 150-200 & $-0.21 \pm 0.11$ & $-0.9 \pm 0.5$ \\
\hline
\hline \hline
\end{tabular}
\end{center}
\end{table}

\begin{table}
\footnotesize
\caption{Count rates of different energy bands for ObsID P0114661003. C1 are the observed count rates, and C2 are the net count rate corrected for background.}
\label{table:2}
\medskip
\begin{center}
\begin{tabular}{c c c c}
\hline \hline
Telescope & Energy band & C1 & C2 \\
          & (keV)       & (cts~s$^{-1}$) & (cts~s$^{-1}$) \\
\hline
   & 1--2.6   & $621.7 \pm 0.4$ & $617.1 \pm 0.6$ \\
LE & 2.6--4.8 & $302.0 \pm 0.3$ & $299.3 \pm 0.4$ \\
   & 4.8--7   & $147.4 \pm 0.2$ & $145.7 \pm 0.2$ \\
   & 7--11    & $79.8 \pm 0.2$  & $73.8  \pm 0.5$ \\
\hline
   & 7--11 & $197.9 \pm 0.2$ & $191.3 \pm 0.5$ \\
ME & 11-23 & $372.6 \pm 0.2$ & $357.9 \pm 1.1$ \\
   & 23-35 & $172.2 \pm 0.2$ & $150.3 \pm 1.5$ \\
\hline
   & 25-35   & $950.2 \pm 0.4$ & $849 \pm 7$ \\
   & 35-48   & $829.4 \pm 0.4$ & $773 \pm 4$ \\
HE & 48-67   & $637.1 \pm 0.3$ & $542 \pm 7$ \\
   & 67-100  & $457.7 \pm 0.3$ & $397 \pm 4$ \\
   & 100-150 & $189.5 \pm 0.2$ & $138 \pm 4$ \\
   & 150-200 & $106.5 \pm 0.1$ & $27 \pm 6$ \\
\hline
\hline \hline
\end{tabular}
\end{center}
\end{table}

\begin{table}
\footnotesize
\caption{General properties of the three typical observations used in Extended Data Fig. 3. $T$ is the net exposure, and $f_{\rm LFQPO}$ is the centroid frequency of the LFQPO.}
\label{table:3}
\medskip
\begin{center}
\begin{tabular}{c c c c c c}
\hline \hline
ObsID & Observed date & $T_{\rm LE}$   & $T_{\rm ME}$  & $T_{\rm HE}$ & $f_{\rm LFQPO}$ \\
      & (MJD)         & (ks) & (ks) & (ks) & (Hz) \\
\hline
P0114661003 & 58200 & 34.8 & 61.0 & 61.0 & $0.038 \pm 0.001$ \\
P0114661036 & 58241 & 8.8  & 14.8 & 14.8 & $0.298 \pm 0.001$ \\
P0114661078 & 58297 & 8.0  & 19.6 & 24.7 & $0.424 \pm 0.002$ \\
\hline
\hline \hline
\end{tabular}
\end{center}
\end{table}

\end{document}